\documentclass[dvips]{article}
\usepackage{icrctc07}

\title{Analysis of Flash ADC Data With VERITAS}
\shorttitle{VERITAS Flash ADC}
\authors{P. Cogan$^1$ for the VERITAS Collaboration$^2$.}
\shortauthors{P. Cogan and et al}
\afiliations{$^1$McGill University, 3600 University Street, Montreal, QC H3A 2T8, Canada \\ 
$^2$For full author list see G. Maier, "Status and Performance of VERITAS", these proceedings}
\email{coganp@hep.physics.mcgill.ca}

\abstract{VERITAS employs a 12m segmented mirror and pixellated
photomultiplier tube camera to detect the brief pulse of Cherenkov
radiation produced by the extensive air shower initiated by a cosmic
high-energy gamma ray. The VERITAS data acquisition system consists of
a 500 Mega-Sample-Per-Second custom-built flash ADC system, which
samples the Cherenkov light pulse every 2 nanoseconds. The integrated
charge in each flash ADC channel is proportional to the amount of
Cherenkov light incident on the corresponding photomultiplier
tube. Accurate reconstruction of the integrated charge is required for
accurate energy estimation and spectral reconstruction. A reliable
calculation of the integrated charge at low intensities can lead to a
reduction in the energy threshold of the system, and an increase in
sensitivity. This paper investigates and compares several approaches
for evaluating the integrated charge. The Cherenkov pulse timing
information in the flash ADC readout has the potential to assist in
background rejection techniques. Various methods for extracting the
timing information are investigated and excellent timing resolution is
achieved.}

\begin{document}
\maketitle

\section{Introduction}
There are many methods \cite{holder05} which can be used to evaluate the digitised Cherenkov signal produced by the VERITAS FADC \cite{buckley03} system. In this paper, five such methods (referred to as \emph{trace evaluators}) are described and the charge integration characteristics of each method compared using a Monte-Carlo simulated photon data set. This study could aid accurate reconstruction of low-intensity events, which is one of the most challenging aspects of the analysis of Cherenkov telescope data. Laser \cite{hanna07} calibration data are used to compare the timing resolution inherent to each trace evaluator, and a digital processing scheme which can further enhance the timing resolution is introduced. These methods have been developed and implemented with VEGAS \cite{cogan07a}.

\section{Methods}

In this section each trace evaluator will be described and the manner in which the integrated charge and pulse arrival time is calculated is discussed. The integrated charge is defined as the sum of the trace in digital counts over some integration window. The pulse arrival time (hereafter $T_0$) is defined as the time at which the pulse reaches $50\%$ of its absolute maximum.

The first method is the \emph{simple-window} trace evaluator which assumes a-priori knowledge of the location of the Cherenkov pulse in the readout window. The second method is the \emph{dynamic-window} trace evaluator which improves on charge integration by sliding an integration window along the readout window to seek the Cherenkov pulse. The first two evaluators only calculate $T_0$ to the nearest sample. The third method is the \emph{linear-interpolation} trace evaluator. This is not significantly different in terms of charge integration, but substantially improves on the calculation of $T_0$. The fourth method is the \emph{trace-fit} trace evaluator which fits the following function to each trace:

\begin{equation}
q(t) = \left\{ \begin{array}{rcl}
    q_{0}\exp{\frac{ -{\left( t - t_{0} \right) }^2}{2\sigma^2}} & \mbox{for} & t \le t_0 \\
    &&\\
    q_{0}\exp{\frac{ -{\left( t - t_{0} \right) }^2}{2\sigma^2 + \alpha\left(t-t_{0}\right)}} & \mbox{for} & t > t_0 
\end{array}\right.
\label{traceFit}
\end{equation}

In this equation, $q(t)$ is the FADC charge at time $t$, $t_0$ is the
time of trace maximum, $q_0$ is the trace amplitude at $t=t_0$, and
$\sigma$ and $\alpha$ are parameters describing the shape of the
trace. This fit function essentially has an asymmetric-Gaussian shape
and improves the calculation of $T_0$ over the \emph{simple-window}
method. The fifth method is the \emph{matched-filter} trace evaluator
which uses a digital filter based on the assumed shape of the FADC
pulse to integrate the charge. The \emph{matched-filter} trace
evaluator is a somewhat more sophisticated than the other methods,
thus it is described here in more detail.

A \emph{matched filter} is so called because its shape is defined by
the expected form of the received data. The \emph{matched filter}'s
pulse shape is a time-reversed version of the expected pulse
shape. Thus for an expected pulse shape $h(t)$, the ideal
\emph{matched-filter} $h_m(t)$ is

\begin{equation}
\label{eqn:mf1}
h_m(t)=h(T-t)
\end{equation}

\noindent for $0\leq t\leq T$ where $T$ corresponds to the end of the
trace. The output from a filtering application is calculated by a
convolution of the input with the filter

\begin{equation}
\label{eqn:mf2}
y(t)=\int_0^Tr(t)h_m(T-t)\:dt
\end{equation}

\noindent which reduces to the cross correlation of $r(t)$ and $h(t)$
with zero lag.

\begin{equation}
\label{eqn:mf4}
y(t)=\int_0^Tr(t)h(t)\:dt
\end{equation}

\begin{figure}
\begin{center}
\includegraphics [width=0.48\textwidth]{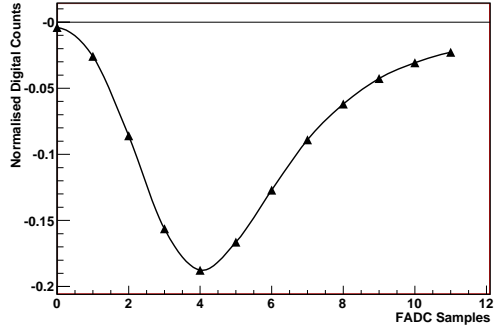}
\end{center}
\caption{Normalised subset of FADC trace used to construct a \emph{matched filter}.}\label{mf}
\end{figure}

In order to construct the matched filter a standard laser calibration
run is used which is normally used to flat field the camera. For each
event, and for each channel, a section of the laser pulse is
extracted, and aligned relative to some predetermined point. This
extracted pulse is summed for all events for each channel. The summed
trace is normalised (Figure \ref{mf}), and Fourier transformed. The
filter is applied to the FADC data by multiplying the Fourier
transform of the FADC trace (denoted $S(\omega)$) with the conjugate
of the filter transform, $\overline{H}(\omega)$, and then applying an
inverse Fourier transform

\begin{equation}
y(t) = \mathcal{F}^{-1}\left[S(\omega) \times \overline{H}(\omega)\right]
\end{equation}

\noindent which yields the cross correlation function $y(t)$. The
maximum of the cross correlation is proportional to the integrated
charge of the FADC trace. In order to establish the
constant, a series of special laser calibration runs is taken with
continuously increasing laser intensity. The integrated charge as
measured using the \emph{dynamic-window} trace evaluator is compared
to the output of the \emph{matched-filter} trace evaluator and used to
establish the constant \cite{cogan07b}.

When analysing data, the charge from a trace is derived by applying
the \emph{matched-filter} trace evaluator, and multiplying the output
by the appropriate constant for that channel. The pulse arrival time is
determined by the location of the maximum of the cross correlation,
thus the arrival time can only be determined to the nearest FADC
sample (much like the \emph{simple-window} trace evaluator).

\section{Integral Charge Evaluation}

\begin{figure}
\begin{center}
\includegraphics [width=0.48\textwidth]{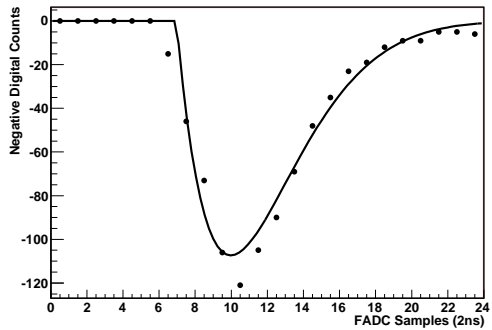}
\end{center}
\caption{Comparison of real and simulated FADC trace. The real trace is indicated by the points.}\label{simCompare}
\end{figure}

In order to examine the charge evaluation quality of each trace
evaluator, a data set of photon impacts on the camera is simulated. The
arrival time of the photons is assumed to be Gaussian with an RMS of 2
ns. The simulation is performed using GrISUDet \cite{duke}. A
comparison of an FADC trace simulated in this way with a real trace
from a laser calibration run is shown in Figure \ref{simCompare}.

\begin{figure}
\begin{center}
\includegraphics [width=0.48\textwidth]{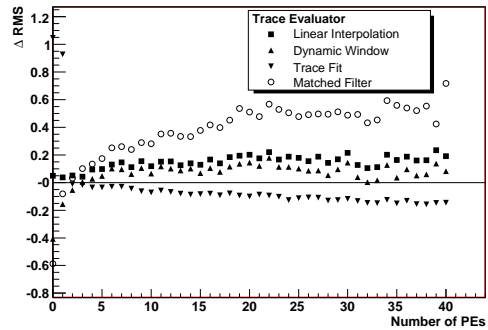}
\end{center}
\caption{Comparison of charge resolution relative to the
\emph{simple-window} trace evaluator.}\label{chargeResCompare}
\end{figure}

The simulated data set is divided into subsets such that each subset
only has events with a certain number of photoelectrons. Data sets
with from one to thirty photoelectrons are generated in this way. This
allows the charge reconstruction as a function of the known number of
photoelectrons to be evaluated. For each trace evaluator, a
distribution of integrated charges (in digital counts) is generated
for each photoelectron multiplicity. In terms of charge evaluation,
the quality of the trace evaluator is determined by the RMS of the
distribution of integrated charges for a constant input. The
difference in the RMS of the \emph{simple-window} trace evaluator, and
each other trace evaluator as a function of the number of
photoelectrons is shown in Figure \ref{chargeResCompare}. Thus, the
RMS of the \emph{simple-window} trace evaluator is used as a baseline
against which the other trace evaluators can be compared. For small
pulses ( $<$ 5 photoelectrons), the \emph{matched-filter} trace
evaluator provides the smallest RMS, however the RMS quickly increases
with the number of photoelectrons. This is to be expected as although
small pulses are dominated by noise, the matched filter is is able to
pick out the signal from the trace. Conversely, the \emph{trace-fit}
trace evaluator gives a very large RMS for small pulses. This is
attributed to ill-fitting of small, poorly-defined pulses. At
approximately four photoelectrons, all the trace evaluators yield
similar results. Beyond that, the \emph{trace-fit} trace evaluator is
superior, and only the \emph{matched-filter} trace evaluator is
significantly worse.

\section{Trace Resampling}

\begin{figure}
\begin{center}
\includegraphics [width=0.48\textwidth]{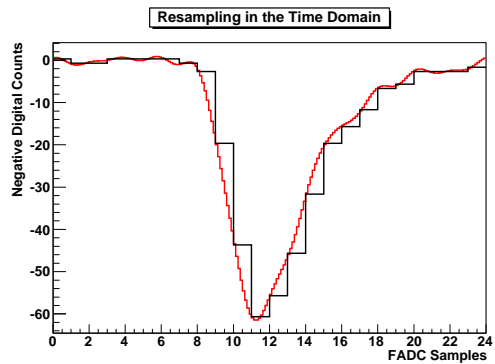}
\end{center}
\caption{Comparison of original and resampled FADC trace.}\label{buildResample}
\end{figure}

One tool commonly used in digital signal processing is
\emph{resampling} in the time domain. The resampling is achieved by
applying a Fourier transform to the FADC trace, \emph{zero-padding} in
the frequency domain, and applying an inverse Fourier transform. Zero
padding in the frequency domain is achieved by simply \emph{adding}
zeros to the end of the Fourier-transformed trace. This has the effect
of setting higher frequencies to have zero amplitude. The
inverse Fourier transform results in a trace which has been resampled
in the time domain. Figure \ref{buildResample} shows a comparison
between a raw and resampled trace. 

\section{Timing Resolution}

The timing resolution is determined by how well the arrival time of an
asynchronous laser flash incident on the camera plane can be
measured. 

\begin{figure}
  \begin{center}
\includegraphics [width=0.48\textwidth]{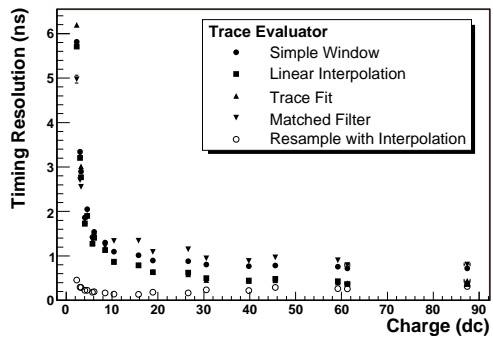}
  \end{center}
  \caption{Timing resolution as a function of FADC trace size.}
    \label{timingRes}
\end{figure}

The timing resolution is defined as the width of a Gaussian function
fit to the distribution of measured differences between event arrival
and channel arrival time for a series of laser pulses for each
channel. The event arrival time is defined as the \emph{average}
arrival time of all the channels in the camera. The timing resolution
as a function of integrated charge is shown for all methods in Figure
\ref{timingRes}. As expected, the timing resolution improves as a
function of pulse size, as for larger pulses the time structure is
better defined.  Above an integrated charge of $30\:\mathrm{dc}$, the
timing resolution is roughly linear as a function of charge. The
timing resolution for pulses greater than $30\:\mathrm{dc}$ is shown
in Table \ref{table}.

\begin{table}[htbp]
  \begin{center}
    \begin{tabular}{ |l|c|c| }
      \hline
      Evaluator                   & Resolution (ns) & Error (ps) \\ \hline
      Simple Window               & 0.77            &  3.2       \\ \hline
      Linear Interpolation        & 0.45            &  2.9       \\ \hline
      Trace Fit                   & 0.46            &  2.9       \\ \hline
      Matched Filter              & 0.91            &  2.7       \\ \hline
      Resampling                  & 0.22            & 10.9       \\ \hline
    \end{tabular}
  \end{center}
  \caption{Timing resolution for pulses greater than $30\:\mathrm{dc}$.}
  \label{table}
\end{table}

\noindent As expected, the \emph{linear-interpolation} trace evaluator
has excellent timing resolution for all trace sizes. The
\emph{matched-filter} trace evaluator is excellent for small traces,
however as pulse arrival times can only be calculated to the nearest
FADC sample, it is not as good for large pulses. The \emph{trace-fit}
trace evaluator has poor timing resolution for very small pulses -
mirroring the effect seen with the study of charge resolution,
indicating that the fit function is not suited to small pulses. For
large pulses the \emph{trace-fit} evaluator has a superior
resolution. However, the best timing resolution is achieved using a
combination of the resampling technique and the
\emph{linear-interpolation} trace evaluator. Together, these fast
robust techniques provide a timing resolution of just
$0.22\:\mathrm{ns}$ with these data.

\section{Conclusions}

Five trace evaluation techniques and a digital filtering technique
have been described and compared using real and simulated data. The
\emph{matched-filter} technique holds promise for the evaluation of
small sub-threshold events. Optimal timing resolution has been
achieved using an FADC resampling technique in concert with the
\emph{linear-interpolation} trace evaluator.

\section{Acknowledgements}

This research is supported by grants from the U.S. Department of Energy, the U.S. National Science Foundation, the Smithsonian Institution, by NSERC in Canada, by PPARC in the UK and by Science Foundation Ireland.

\bibliography{icrc0575}

\bibliographystyle{plain}

\end{document}